\documentclass[reprint,showpacs,preprintnumbers,amsmath,amssymb, aip]{revtex4-1}
\usepackage{graphicx,wasysym}% Include figure files
\usepackage{dcolumn}% Align table columns on decimal point
\usepackage{bm}% bold math
\usepackage{natbib}
\usepackage{tabulary}
\usepackage{natmove}

\begin{document}

\preprint{1}

\title{Origin of reversible photo-induced phase separation in hybrid perovskites}% Force line breaks with \\

\author{Connor G. Bischak}
\affiliation{Department of Chemistry, University of California, Berkeley, California 94720, United States}
\author{Craig L. Hetherington}
\affiliation{Department of Chemistry, University of California, Berkeley, California 94720, United States}
\affiliation{Molecular Biophysics and Integrative Bioimaging Division, Lawrence Berkeley National Laboratory, Berkeley, California, 94720, United States}
\author{Hao Wu}
\affiliation{Department of Chemistry, University of California, Berkeley, California 94720, United States}
\affiliation{Current address: Department of Chemistry and Chemical Biology, Harvard University, Cambridge, MA 02138, United States}
\author{Shaul Aloni}
\affiliation{Materials Science Division, Lawrence Berkeley National Laboratory, Berkeley, California, 94720, United States}
\affiliation{Molecular Foundry, Lawrence Berkeley National Laboratory, Berkeley, California, 94720, United States}
\author{D. Frank Ogletree}
\affiliation{Materials Science Division, Lawrence Berkeley National Laboratory, Berkeley, California, 94720, United States}
\affiliation{Molecular Foundry, Lawrence Berkeley National Laboratory, Berkeley, California, 94720, United States}
\author{David T. Limmer}
\affiliation{Department of Chemistry, University of California, Berkeley, California 94720, United States}
\affiliation{Kavli Energy NanoScience Institute, Berkeley, California 94720, United States}
\author{Naomi S. Ginsberg}\email{nsginsberg@berkeley.edu}
\affiliation{Department of Chemistry, University of California, Berkeley, California 94720, United States}
\affiliation{Molecular Biophysics and Integrative Bioimaging Division, Lawrence Berkeley National Laboratory, Berkeley, California, 94720, United States}
\affiliation{Materials Science Division, Lawrence Berkeley National Laboratory, Berkeley, California, 94720, United States}
\affiliation{Kavli Energy NanoScience Institute, Berkeley, California 94720, United States}
\affiliation{Department of Physics, University of California, Berkeley, California 94720, United States}

\date{\today}

\maketitle
\textbf{Nonequilibrium processes occurring in functional materials can significantly impact device efficiencies and are often difficult to characterize due to the broad range of length and time scales involved. In particular, mixed halide hybrid perovskites are promising for optoelectronics, yet the halides reversibly phase separate when photo-excited, significantly altering device performance. By combining nanoscale imaging and multiscale modeling, we elucidate the mechanism underlying this phenomenon, demonstrating that local strain induced by photo-generated polarons promotes halide phase separation and leads to nucleation of light-stabilized iodide-rich clusters. This effect relies on the unique electromechanical properties of hybrid materials, characteristic of neither their organic nor inorganic constituents alone. Exploiting photo-induced phase separation and other nonequilibrium phenomena in hybrid materials, generally, could enable new opportunities for expanding the functional applications in sensing, photoswitching, optical memory, and energy storage.}

Photovoltaic and light-emitting devices typically operate under conditions far from equilibrium. As such, elucidating the response of functional materials to nonequilibrium driving forces is vital to understanding their fundamental properties and to determining their suitability for device applications. In particular, photo-induced dynamic processes are of major importance to the performance of hybrid perovskite-based devices \cite{hoke2015reversible,gottesman2016perovskites,zhang2016photo}. Hybrid perovskites are low-cost, solution processable materials that are promising for many device applications, including photovoltaics \cite{lee2012efficient,stranks2013electron,burschka2013sequential,zhou2014interface,nie2015high,mcmeekin2016mixed} and light-emitting diodes (LEDs)\cite{tan2014bright}. The high device efficiencies have been attributed to their high brightness, long charge carrier migration lengths \cite{stranks2013electron,dong2015electron}, and tolerance of structural defects \cite{de2015impact,bischak2015heterogeneous}. The chemical formula of hybrid perovskites is APbX$_3$, where A is an organic cation, typically methylammonium ((CH$_3$NH$_3^+$, MA) or formamidinium (HC(NH$_2$)$_2^+$, FA), and X is either iodide, bromide, chloride, or iodide/bromide or bromide/chloride mixtures. By varying the halide ratios in hybrid perovskites, the bandgap can be tuned across the visible spectrum \cite{hoke2015reversible,noh2013chemical,sadhanala2015blue}. Precise control of the bandgap presents promising opportunities for color tuning perovskite-based LEDs and lasers, and for incorporating hybrid perovskites in tandem solar cells \cite{mcmeekin2016mixed,bailie2015semi}. Light-induced effects, however, restrict the practical use of mixed halide hybrid perovskites \cite{hoke2015reversible,beal2016cesium,jaffe2016high}.  Photoluminescence (PL) and X-ray diffraction (XRD) measurements suggest that MAPb(I$_x$Br$_{1-x}$)$_3$ (0.1 $< x <$ 0.8) undergoes reversible phase separation into iodide-rich and bromide-rich regions when photo-excited \cite{hoke2015reversible}. Such demixing is detrimental to photovoltaic performance, since it leads to charge carrier trapping in the iodide-rich regions. Determining the microscopic mechanism behind phase separation is essential for furthering approaches to mitigate adverse photo-induced effects in devices and should expand the range of their functional applications into areas such as optical memory storage and sensing \cite{lin2015low,sutherland2015sensitive}. Unfortunately, the microscopic mechanism behind this effect has been elusive because of the multiple length and time scales involved in characterizing the behavior both experimentally and theoretically.

We find that these novel photo-induced processes arise because hybrid materials, such as perovskites, metal-organic frameworks, and nanocrystal superlattices, have physical properties that differ substantially from both traditional inorganic and organic semiconductors. In particular, hybrid perovskites have elastic moduli that fall in between those of pure inorganic and organic solids \cite{rakita2015mechanical,sun2015mechanical}, aiding in their tolerance of structural defects while maintaining long range order. Hybrid perovskites also have a high static dielectric constant, resulting in small exciton binding energies and strong electron-phonon coupling \cite{onoda1992dielectric,johnston2015hybrid}. These macroscopic properties reflect the broad range of timescales characterizing molecular motion in these materials. For example, motions associated with freely rotating organic cations and facile halide migration have been implicated in the high polarizability of the material \cite{mattoni2015methylammonium, leguy2015dynamics} and in hysteresis upon charging \cite{snaith2014anomalous, unger2014hysteresis}. Understanding the interplay between, and molecular origin of, the mechanical and electronic properties of these hybrid materials is critical for maximizing their utility for frontier device applications and for the development of new high performance functional materials. 

Here we demonstrate that photo-induced phase separation in mixed halide hybrid perovskites is mediated through strain induced at the locations of polarons-- photogenerated charge carriers and their accompanying lattice distortions. We combine cathodoluminescence (CL) imaging, a powerful tool for characterizing the nanoscale optical properties of hybrid perovskite films \cite{bischak2015heterogeneous,dou2015atomically, xiao2015mechanisms}, and multiscale modeling, a set of techniques to bridge molecular and mesoscopic scales, to observe and explain the dynamic process of photo-induced phase separation. After prolonged illumination, small clusters enriched in one halide species are observed to localize near grain boundaries, which is consistent with the effects of polaronic strain in both our molecular and mean-field phenomenological models and the associated phase diagram that we construct. The transient dynamics of cluster formation are characterized by an initial latency and stochastic fluctuations in their formation. The process of cluster formation is captured with the phenomenological theory at intermediate electron-phonon coupling and consists of photogenerated polarons finding, stabilizing, and subsequently becoming trapped in halide composition fluctuations. Additional imaging and calculations validate the proposed requirement that polarons reside specifically within the lower bandgap phase-separated regions to ensure their stability and suggest a path toward new optical memory or detection applications.

\begin{figure}[t]
\begin{center}
\includegraphics[width=8.5cm]{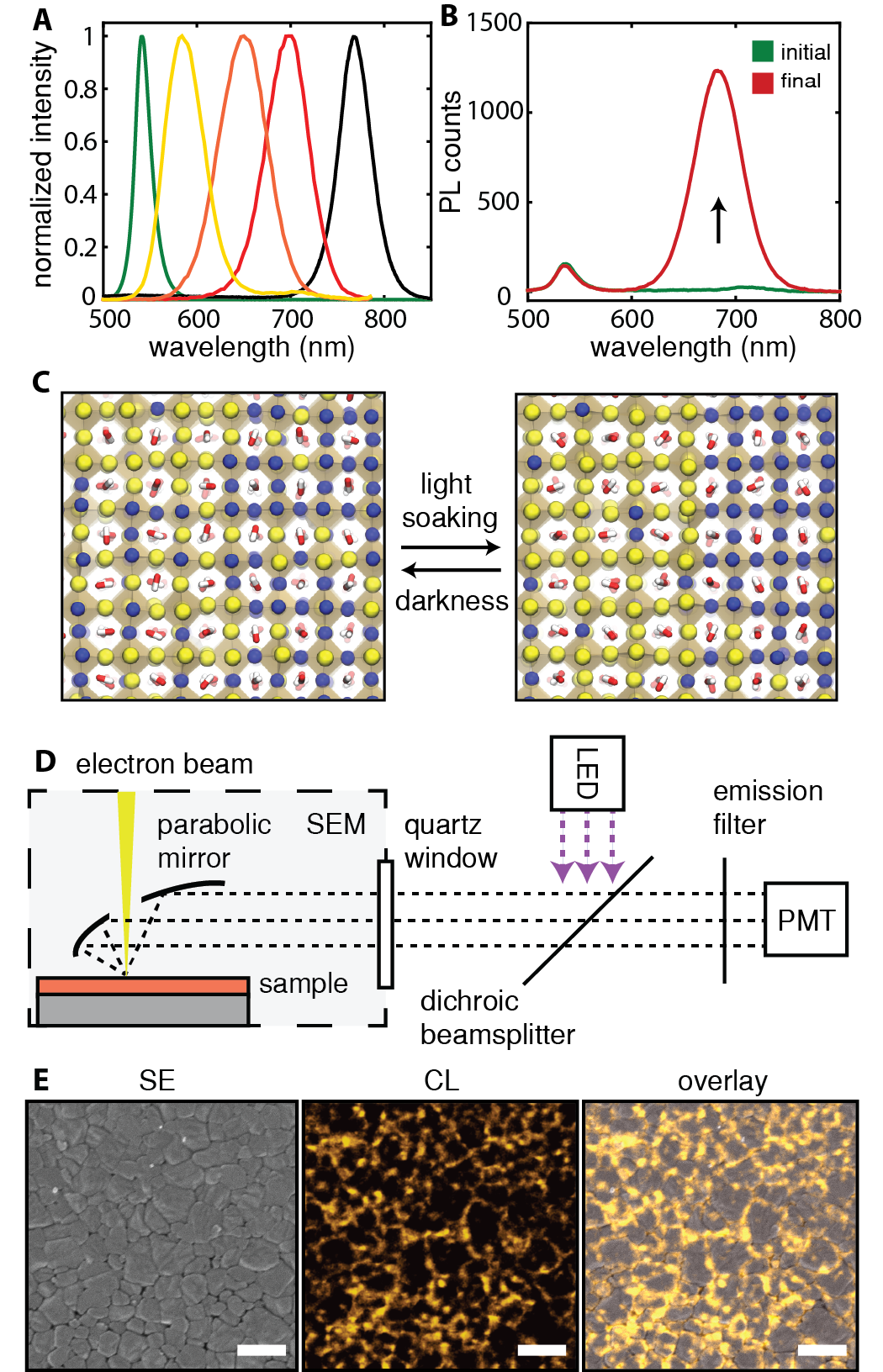}
\caption{{\bf Band gap tunability, PL spectra before and after light soaking, CL experimental set-up, and CL/SE images at steady state.} (A) PL spectra of MAPb(I$_x$Br$_{1-x}$)$_3$ films with varying $x$. (B) PL spectra of MAPb(I$_{0.1}$Br$_{0.9}$)$_3$ before (green) and after (red) light soaking for 5 min. (C) Schematic of phase separation and reversibility in MAPb(I$_x$Br$_{1-x}$)$_3$ where yellow and blue spheres represent I$^-$ and Br$^-$, respectively. (D) Schematic of CL acquisition and light soaking. (E) SE, CL, and SE/CL overlay after light soaking for 5 min at 100 mW/cm$^2$. The scale bar is 2 $\mu$m.}
\label{Fi:1}
\end{center} 
\end{figure}

To observe photo-induced phase separation, we fabricated mixed halide hybrid perovskite films for CL imaging. We made a series of films with varying iodide:bromide ratios (Figure 1a) and ultimately selected MAPb(I$_{0.1}$Br$_{0.9}$)$_3$ for further investigation because of its spectrally distinct iodide-rich and bromide-rich regions. When initially photo-exciting the MAPb(I$_{0.1}$Br$_{0.9}$)$_3$ film, we observe a single emission peak at 540 nm, corresponding to well-mixed halides. Upon continued illumination, a second spectral feature appears at 690 nm (Figure 1b), which corresponds to iodide-rich content and is thus indicative of phase separation (Figure 1c). When placed in the dark, the halides remix and the red emission peak disappears on a similar time scale. We illuminate the sample with a 405 nm LED for light soaking inside a scanning electron microscope (SEM) fitted with CL collection capabilities (Figure 1d) and then scan the electron beam to probe photo-induced iodide-rich cluster formation at the nanoscale. The electron beam by itself cannot induce phase separation (see supplementary text). Figure 1e shows a secondary electron (SE) image, CL image, and a SE/CL overlay collected after 5 min illumination at 100 mW/cm$^2$, demonstrating that the iodide-rich regions localize to grain boundaries in steady state. Although the size of the bright spots in the CL image results from a convolution of iodide cluster size and a large carrier migration length, individual clusters are still visible within each grain.

\begin{figure*}[t]
\begin{center}
\includegraphics[width=15cm]{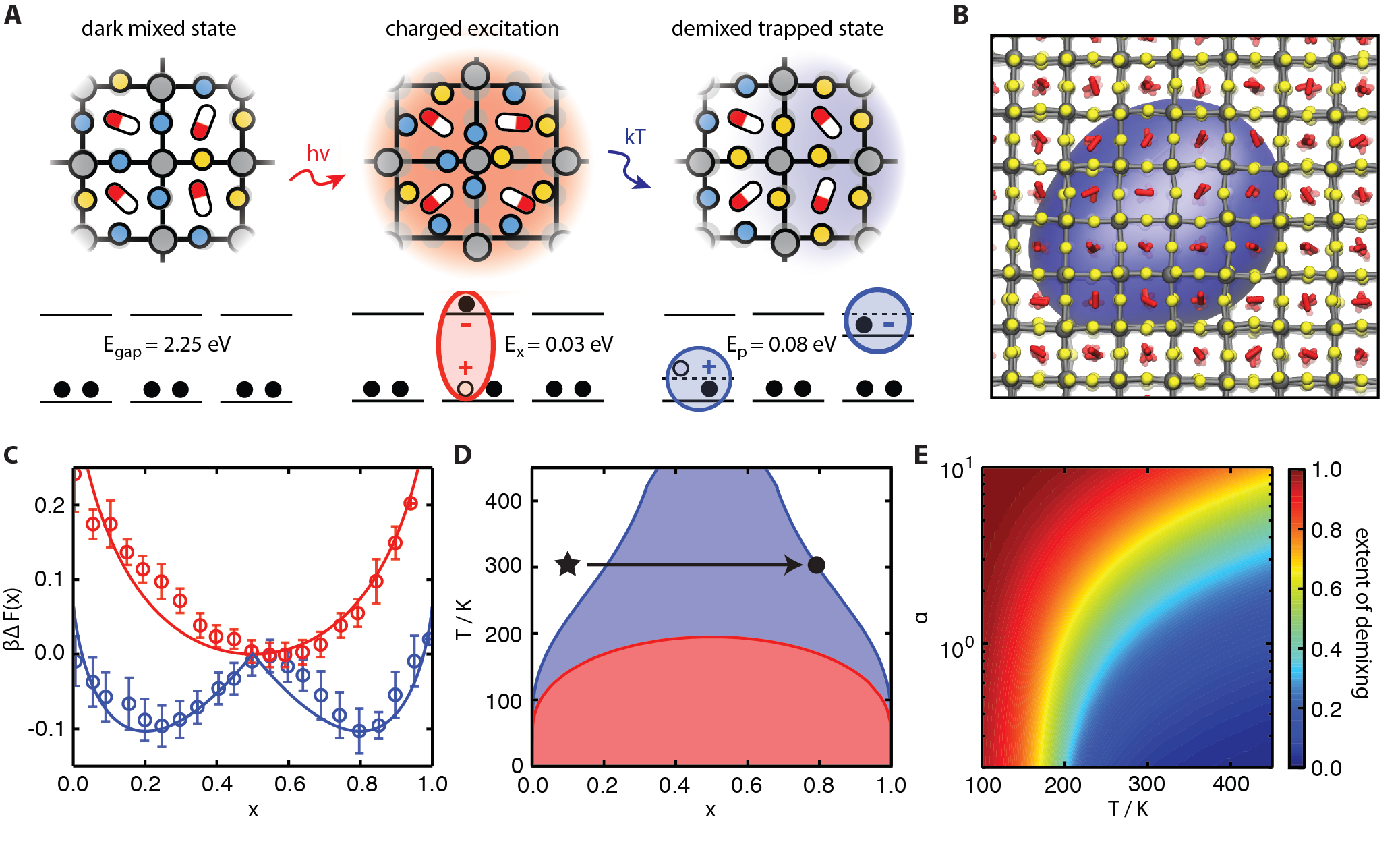}
\caption{{\bf Steady-state stabilities and dynamic mechanism for photo-induced phase separation.} (A) Photo-induced polaron trapping and associated energy scales associated with phase separation. Color scheme is the same as in Figure 1. (B) Snapshot of the 99$\%$ isosurface of excess charge density taken from the molecular dynamics (MD) simulation. (C) Free energies per unit cell for MAPb(I$_x$Br$_{1-x}$)$_3$ with varying composition, in the ground (red) and photo-excited (blue) states, computed from MD simulations (circles) and mean field theory (solid lines). (D) Mean field theory temperature-composition phase diagram in the ground (red) and photo-excited state (blue) with the path through the phase diagram from initial state (star) to demixed state (circle) observed experimentally. Areas beneath the red and blue coexistence curves indicate demixed states. (E) The extent of demixing (i.e., purity of demixed regions found by tracing the blue coexistence curve in D) as a function of electron-phonon coupling and temperature computed from mean field theory in the photo-excited state.}
\label{Fi:1}
\end{center} 
\end{figure*}
	
Molecular simulations clarify that photo-induced phase separation is a consequence of charged excitations that generate sufficient lattice strain to destabilize the solid solution and favor demixing. Using a classical point charge model \cite{mattoni2015methylammonium, lewis1985potential} we find that mixed I/Br perovskites undergo demixing transitions as a function of temperature with a critical temperature of 190 K (Figure S1), consistent with the well-mixed films made experimentally at room temperature. By analyzing the relative energetic contributions to the heat of mixing, we find that elastic effects from lattice mismatch are much larger than specific chemical interactions, which leads to a demixing transition that depends strongly on strain (Figure S2). As illustrated in Figure 2a, upon light absorption weakly bound electron-hole pairs, with binding energies $E_\mathrm{x} = $0.03 $e$V\cite{johnston2015hybrid},  rapidly dissociate, easily creating free charges. These charges deform the surrounding lattice through electron-phonon coupling, which is expected to be significant given the ionic nature of the material. Using free energy calculations and path integral molecular dynamics of a pseudopotential-based model on an excess charge\cite{kuharski1988molecular}, shown in Figure 2b and Movie S1, we find that high spatial overlap between the lattice and a single-charge density distribution generates sufficient strain to drive local phase separation at room temperature. Figure 2c shows the free energy with and without the excess charge as a function of composition, $\beta \Delta F(x)$,  within a volume characteristic of the charge distribution's extent. This charge and the lattice deformation field that surrounds it together form a polaron that we predict to have an average size of 8 nm and binding energy $E_\mathrm{p} = $0.08 $e$V (Figure 2a). As discussed below, the lower bandgap of the iodide-enriched phase energetically stabilizes and spatially traps the polaron. 

\begin{figure*}[t]
\begin{center}
\includegraphics[width=15cm]{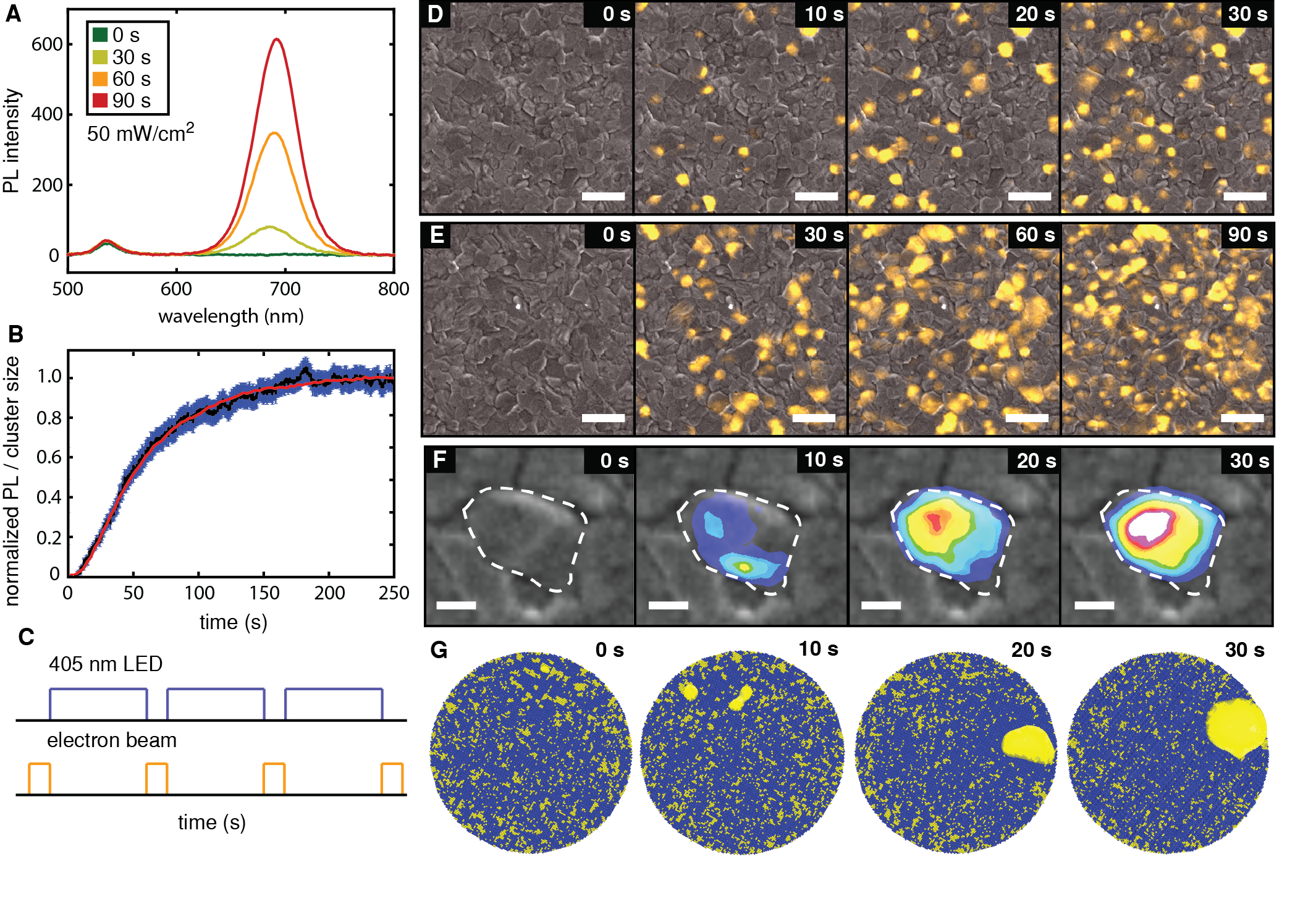}
\caption{{\bf Formation and evolution of iodide-rich clusters.} (A) PL spectra after different light soaking times at 50 mW/cm$^2$. (B) Normalized PL intensity versus time (red) and normalized simulated cluster size versus time (black) with standard error values (blue). (C) Duty cycle for CL image series. CL image series with (D) 10 sec and (E) 30 sec of light soaking between each CL image. The scale bars are 2 $\mu$m. (F) CL image series of a single domain with 10 sec of light soaking between each CL image. The color scale indicates iodide-rich CL intensity, which convolves carrier migration length with feature size.The scale bar is 200 nm. (G) A series of snapshots from a cluster formation simulation of a 100 nm region with iodide-rich regions in yellow and bromide-rich regions in blue.}
\label{Fi:1}
\end{center} 
\end{figure*}

To relate these findings to the experimental observations, we have distilled them into a simple analytical theory. Modeling the phase separation with a Landau-Ginzburg hamiltonian with linear coupling between strain and composition fields\cite{cahn1962spinodal}, and applying a semi-classical description of the excess charge\cite{kalosakas1998polaron}, we parameterize and evaluate a theory for photo-induced phase separation (see supplementary text). As shown in Figure 2c, this model is capable of describing the underlying free energy surfaces predicted from molecular dynamics simulations. Within a mean-field approximation, we determine the full temperature-composition phase diagram for both ground- and photo-excited states (Figure 2d). We confirm experimentally that varying the film temperature over a range of 50 K is insufficient to induce demixing, but does increase the demixing rate by increasing halide mobility (Figure S3). We also map the degree of demixing as a function of temperature and electron-phonon coupling, $\alpha$, in the photo-excited state (Figure 2e). Previous PL studies confirm that the stable iodide-rich cluster phase is at approximately MAPb(I$_{0.8}$Br$_{0.2}$)$_3$ \cite{hoke2015reversible}, which agrees with our predicted phase diagram and PL measurements. 

Having established a model for the thermodynamics of halide phase separation, we focus on the dynamics of the formation and evolution of iodide-rich clusters, which further constrains our model. To monitor the formation of iodide-rich clusters in bulk, we collected a series of PL emission spectra under constant light-soaking at 50 mW/cm$^2$ (Figure 3a) and plot the integrated intensity of iodide-rich emission as a function of time as it grows to a finite value proportional to the illumination intensity (Figure 3b, red). To resolve the emergence and evolution of iodide-rich clusters, we alternate between light soaking and CL imaging (Figure 3c). Figures 3d and 3e (Movie S2 and S3) show overlaid CL and SE image sequences with 10- and 30-s intervals of light soaking, respectively. After some latency, iodide-rich clusters begin to emerge and grow primarily in number. They quickly reach a maximum size, suggesting that cluster nucleation rather than growth limits the rate of phase separation. We also find using CL microscopy that the rate of cluster formation increases with illumination intensity (Figure S7, Movie S4 and S5). 

To gain insight into the microscopic dynamics leading to cluster formation, we examine the photo-induced progression within single domains with both CL (Figure 3f and Figure S8) and simulations (Figure 3g). Simulations of the clustering process are performed by solving our phenomenological theory numerically on a lattice (Movie S6). We observe similar dynamics in both experiment and simulation, such as the formation of both transient and stabilized clusters. An ensemble average of independent simulated clustering events yields a curve that characterizes the clustering process (Figure 3b, black), which agrees with the experimentally observed PL intensity growth in time (red) and its illumination power dependence (Figures S9 and S10). We find that the 5-10 s initial lag time corresponds to a characteristic time scale for a polaron to become trapped in a spontaneous fluctuation of higher iodide concentration, which is limited by the ions' diffusivity rather than the polaron's. During the following 10-90 s, a trapped polaron stabilizes an iodide-rich cluster and accumulates more iodide, and more clusters form within the film. Subsequently, cluster growth stops, as cluster size is limited to the deformation region of the polaron and cluster number by the total number of photogenerated charges. Although the time scales for these dynamics are considerably longer than the lifetime of a single polaron, continuous illumination enables newly generated polarons to replace recombining ones.  By measuring the PL of the iodide-rich clusters at different intensities (Figure S9), which correspond to different steady-state polaron densities, we estimate the size of the clusters to be 8-10 nm in diameter (Figure S11), in agreement with our theoretical prediction. During later times, the clusters migrate to grain boundaries to relieve strain (Figure 1e).

To confirm the validity of our thermodynamic model for photo-induced phase separation, we demonstrate experimentally both the ability of polarons to stabilize iodide-rich clusters and that large electron-phonon coupling is required to cause halide phase separation. For the latter, we show both experimentally and with the phenomenological model that reducing the electron-phonon coupling in the system by replacing MA$^+$ with less polar Cs$^+$ reduces the tendency to phase separate (Figure S4), consistent with the phase diagram in Figure 2e. To demonstrate that the local presence of a polaron stabilizes iodide-rich clusters in the hybrid perovskites, specifically in MAPb(I$_{0.1}$Br$_{0.9}$)$_3$, we first generate iodide-rich clusters with a 405 nm excitation source. We then illuminate the film with 635 nm light, generating charge carriers only in pre-existing iodide-rich clusters. The 635 nm light stabilizes these clusters, but does not generate new ones (Figure S5 and S6), confirming that the continued presence of photo-generated carriers is required for cluster stability. 

The confluence of multiple physical processes that couple disparate length and time scales gives rise to a decidedly peculiar photo-induced phase separated state. For example, because the strain field applied by each polaron is limited in spatial extent, the photo-excited steady state of the system consists of a series of isolated nanoscale clusters rather than the expected large, single iodide-rich domain of a low-temperature equilibrium state in the dark (Figure 3d). Further, because polarons preferentially localize in iodide-rich regions, the composition of the remaining material is neither predicted nor observed (Figure 1b) to deviate from its pre-illumination halide mixture ratio, as though it were still in the dark. Last, because the iodide-rich clusters are limited to nanoscale sizes, their coarsening dynamics do not abide by familiar, universal scaling laws. 

We have established that the unusual interplay between the electronic and mechanical properties in hybrid perovskite materials results in local, excited-state phase behavior that differs substantially from the ground state due to polaronic strain. Photo-induced phase separation is a general phenomenon in hybrid perovskites that has been observed in both polycrystalline films and single crystals, and with different organic cations\cite{hoke2015reversible}. We further find both experimentally and theoretically that it occurs with other halide mixtures, such as in MAPb(Br$_x$Cl$_{1-x}$)$_3$ thin films (Figures S12 and S13). Our study shows that the unique combination of mobile halides, substantial electron-phonon coupling, and long-lived charge carriers is required for photo-induced phase separation. Decreasing defect concentrations to reduce vacancy-mediated halide migration or lowering electron-phonon coupling could significantly reduce photo-induced effects and improve compatibility with device applications at ambient conditions. For instance, it has been shown that Cs-doped FAPb(I$_x$Br$_{1-x}$)$_3$ films are much more phase-stable under illumination than undoped films, likely due to a decrease in the electron-phonon coupling\cite{mcmeekin2016mixed}.

Taking advantage of photo-induced phase separation could provide new opportunities for expanding the functional applications of mixed halide hybrid perovskites. Due to their sensitive spectral photo-response, these materials could be used in sensing, photoswitching, or memory applications. In fact, we have demonstrated a first step towards memory storage by transiently patterning the local halide composition (Figure S6). More broadly, the commonality in the electronic and mechanical properties among a rapidly growing library of new hybrid materials suggests that nonequilibrium processes in these materials are similar and could be exploited for new device applications. The coordinated multiscale methodologies that we have developed could also uncover the nature of other nonequilibrium phenomena in other dynamic materials-- for example in energy storage\cite{li2014current} or electronically correlated materials\cite{schmitt2008transient}- for which their complexity necessitates a multifaceted approach, capable of bridging molecular and mesoscopic length scales.

{\bf Acknowledgements:} CL characterization was supported by a David and Lucile Packard Fellowship for Science and Engineering to N.S.G. CL and PL at the Lawrence Berkeley Lab Molecular Foundry were performed as part of the Molecular Foundry user program, supported by the Office of Science, Office of Basic Energy Sciences, of the U.S. Department of Energy under Contract No. DE-AC02-05CH11231. C.G.B. acknowledges an NSF Graduate Research Fellowship (DGE 1106400) and N.S.G. acknowledges an Alfred P. Sloan Research Fellowship and a Camile Dreyfus Teacher-Scholar Award. We thank Z. Luo for assistance with film deposition. D.T.L was supported initially through the Princeton Center of Theoretical Science.

\bibliography{apssamp}

%\bibliography{ref}
% Produces the bibliography via BibTeX.

\end{document}